\documentclass[a4paper,UKenglish]{lipics}
 
\usepackage{microtype}

\title{Using Automated Theorem Provers to Teach Knowledge Representation in First-Order Logic}

\author{Angelo Kyrilov}
\author{David C. Noelle}
\affil{University of California, Merced\\
  5200 North Lake Road, Merced, CA, 95343, USA\\
  \texttt{\{akyrilov, dnoelle\}@ucmerced.edu}}
\authorrunning{A. Kyrilov and D. C. Noelle} 

\Copyright{Angelo Kyrilov and David Noelle}

\subjclass{K.3 Computers and Education}
\keywords{Automated Assessment, Knowledge Representation, Automated Theorem Provers}

\serieslogo{logo_ttl}
\volumeinfo
  {M. Antonia {Huertas}, Jo\~ao {Marcos}, Mar\'ia {Manzano}, Sophie {Pinchinat}, \\
  Fran\c{c}ois {Schwarzentruber}}
  {5}
  {4th International Conference on Tools for Teaching Logic}
  {1}
  {1}
  {85}
\EventShortName{TTL2015}

\begin{document}

\maketitle

\begin{abstract}
Undergraduate students of artificial intelligence often struggle with representing knowledge as logical sentences. This is a skill that seems to require extensive practice to obtain, suggesting a teaching strategy that involves the assignment of numerous exercises involving the formulation of some bit of knowledge, communicated using a natural language such as English, as a sentence in some logic. The number of such exercises needed to master this skill is far too large to allow typical artificial intelligence course teaching teams to provide prompt feedback on student efforts. Thus, an automated assessment system for such exercises is needed to ensure that students receive an adequate amount of practice, with the rapid delivery of feedback allowing students to identify errors in their understanding and correct them.
This paper describes an automated grading system for knowledge representation exercises using first-order logic. A resolution theorem prover, \textit{Prover9}, is used to check if a student-submitted formula is logically equivalent to a solution provided by the instructor. This system has been used by students enrolled in undergraduate artificial intelligence classes for several years. Use of this teaching tool resulted in a statistically significant improvement on first-order logic knowledge representation questions appearing on the course final examination. This article explains how this system works, provides an analysis of changes in student learning outcomes, and explores potential enhancements of this system, including the possibility of providing rich formative feedback by replacing the resolution theorem prover with a tableaux-based method.
 \end{abstract}

\section{Introduction}
Undergraduate computer science curricula often provide students with opportunities to study artificial intelligence (AI). Courses on AI frequently cover the development of intelligent systems by constructing knowledge bases composed of logical sentences and performing automated reasoning over those sentences. Computer science students regularly have little background in formal logics before attending an AI course, and this makes the learning of logic-based knowledge representation schemes particularly challenging~\cite{bryce2012}.
\noindent
In the Computer Science and Engineering program at the University of California, Merced, the ``Introduction to Artificial Intelligence'' class provides a broad survey of AI methods and topics, including the construction of automated reasoning systems using first-order logic to represent knowledge. This is an upper-division semester-long undergraduate course which is taught annually. Historically, students enrolled in this class have found knowledge representation to be a particularly difficult topic. When asked to translate English sentences into first-order logic, using a specified ontology, as part of a written final examination, their performance has been extremely poor. Students only score about 30\% of the maximum possible credit, on average, when presented with exam questions of this kind.

These low scores are likely the result of a lack of adequate practice with first-order logic. The broad array of material covered in this survey course limits the amount of lecture time available to illustrate the construction of logical formulae, and high enrollments limit the amount of guidance and feedback each student can expect to receive from the teaching team. While student understanding would certainly benefit from extensive practice on knowledge representation exercises, the grading of such exercises is demanding, as there are often many equally correct ways to express a proposition in first-order logic. Thus, given that students require feedback on practice exercises for them to be useful, the number of exercises that could be assigned has been highly restricted by limited human resources.

In order to address this problem, we built an online repository of exercises involving the translation of English sentences into first-order logic, and we designed and implemented an online software tool to automatically assess student solutions to these exercises. By using this tool, students received instant feedback in the form of ``Correct/Incorrect'' judgments, and students who submitted incorrect solutions were allowed to revise and resubmit their answers. There was no limit on the number of resubmissions permitted.

This online educational system was used in our ``Introduction to Artificial Intelligence'' course during the 2012, 2013, and 2014 offerings. We analyzed student performance on final examination knowledge representation questions, and we compared it to the performance of students from previous years, who had no access to our system. We found that students who used our system exhibited significantly improved scores on the first-order logic knowledge representation questions. 

The rest of this paper is organized as follows. Section~\ref{description} describes the automated grading system in detail. Section~\ref{evaluation} provides an assessment of the system in terms of its contribution to student learning outcomes. Section~\ref{future_work} offers ideas for improving the automated grading software by utilizing semantic tableaux methods instead of resolution theorem proving for the production of exercise feedback. Section~\ref{conclusion} contains some concluding remarks.

\section{System Description}
\label{description}

Our goal was to give students much more practice on knowledge representation exercises. We generated a repository of questions in which students were given an English sentence and were asked to translate it into first-order logic. Each question supplied an explicit list of predicates, functions, and constant symbols that students were allowed to use in their answers. A typical example would be: 
\begin{center}
\begin{minipage}{12cm}
Translate the sentence ``\emph{All surgeons are doctors}'', using the following constants: $Doctor$, $Surgeon$, and predicates: $Occupation(x, y)$.
\end{minipage}
\end{center}
A correct solution to this exercise is the formula:
\begin{displaymath}
	\forall x~Occupation(x, Surgeon) \Rightarrow Occupation(x, Doctor)
\end{displaymath}
It is important to note that there are usually multiple correct solutions to exercises of this kind. For example, another correct answer to the question, above, is:
\begin{displaymath}
        \neg ( \exists x~Occupation(x, Surgeon) \wedge \neg Occupation(x, Doctor))
\end{displaymath}
Thus, student submissions could not be assessed by performing a simple string comparison, or the like, with a correct solution provided by the instructor.

Our system does require the instructor to provide a model answer for each exercise, but it does \emph{not} necessarily label submissions that deviate from this model answer as incorrect. Instead, any submitted formula that is found to be logically equivalent to the model answer is recognized as a correct solution to the exercise. We use the \textit{Prover9} automated theorem prover to check for logical equivalence. If $A$ is the model answer and $B$ is the student solution, the solution is labeled as correct if and only if the formula $A \Leftrightarrow B$ is found to be valid. Prover9 is a resolution based automated theorem prover for first-order logic with equality~\cite{prover9}. Prover9 was selected because it is very easy to use and the syntax of its interface is very similar to what students see in lectures.

\begin{figure}[!htbp]
	\begin{center}
		\includegraphics[scale=0.38]{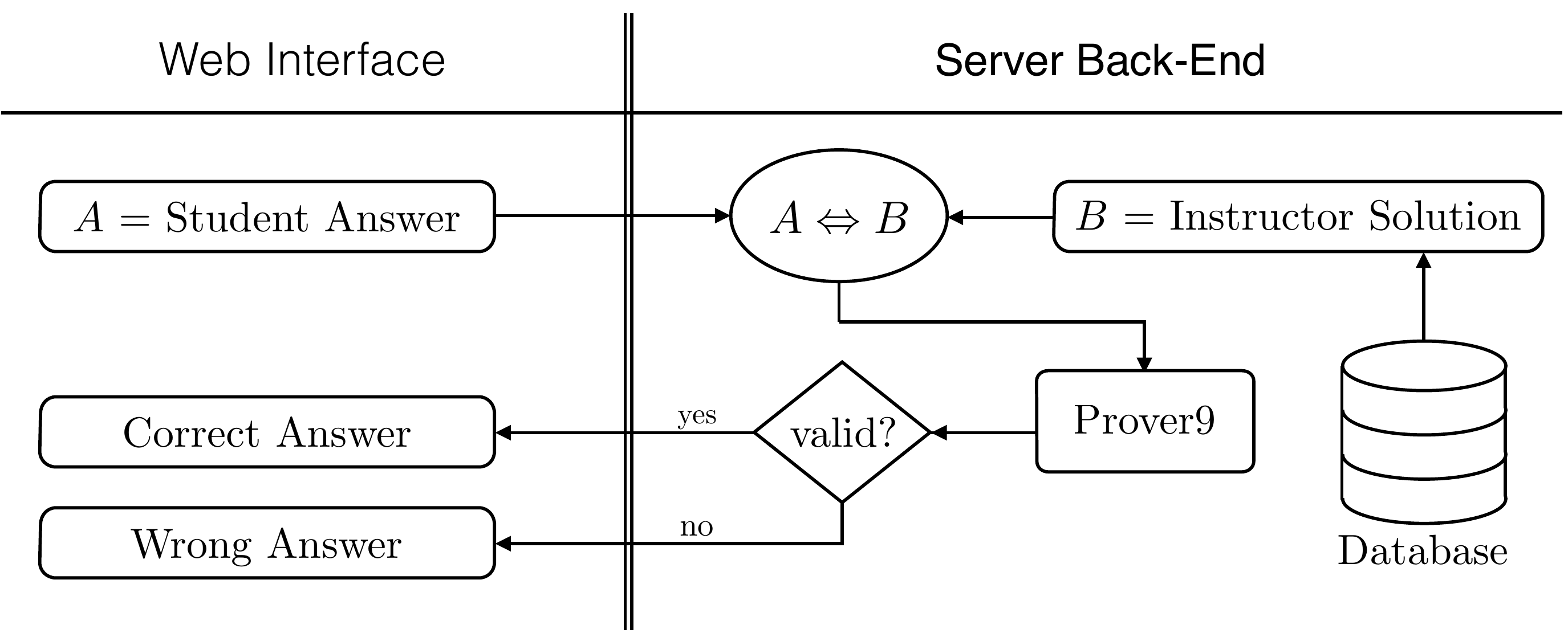}
	\end{center}
	\caption{Components of the Automated Grading System}
	\label{system}
\end{figure}
\noindent
Figure \ref{system} illustrates the automated grading system components, including the web interface and the back-end. When a student submits a solution to an exercise, the model answer is retrieved from the exercise database. Prover9 is used to determine whether the student's solution is logically equivalent to the model answer, and appropriate feedback is immediately sent to the student.

There is a restriction on the amount of time the server is allowed to spend on checking a student's submission. By default, this is set to 5 seconds but it can be adjusted on a per exercise basis. If the time limit is exceeded, an appropriate message is sent to the student informing them that the time limit has been exceeded. While this does not necessarily indicate that the student's answer is incorrect, students are encouraged to revise their solution or talk to an instructor. This takes care of the fact that the prover may run forever due to the undecidability of first-order logic.

\begin{figure}[!htbp]
	\begin{center}
		\includegraphics[scale=0.5]{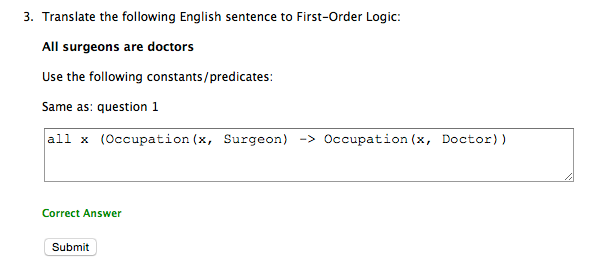}
	\end{center}
	\caption{The student user-interface of the automated grading system}
	\label{interface}
\end{figure}

Figure~\ref{interface} shows the user-interface of the system, which appears in a web browser window. In addition to the question listing, which is what students would see, there is also an administrative interface, allowing instructors to create and assign exercises. 

\section{System Evaluation}
\label{evaluation}

The final examination for the ``Introduction to Artificial Intelligence'' course is a three hour comprehensive written test that covers the full range of AI topics presented during the semester long class. Students complete the exam without access to any textbooks, notes, or other study materials. The final examination contains three questions involving the translation of English sentences into first-order logic, as well as a question asking students to produce a successor-state axiom for a given time-varying predicate~\cite{russell2010}. If the extensive practice afforded by our automated grading system is a benefit to student learning, then we would expect to see higher scores on these particular final examination questions when students made use of our system.

\begin{figure}[!htbp]
	\begin{center}
		\includegraphics[scale=0.38]{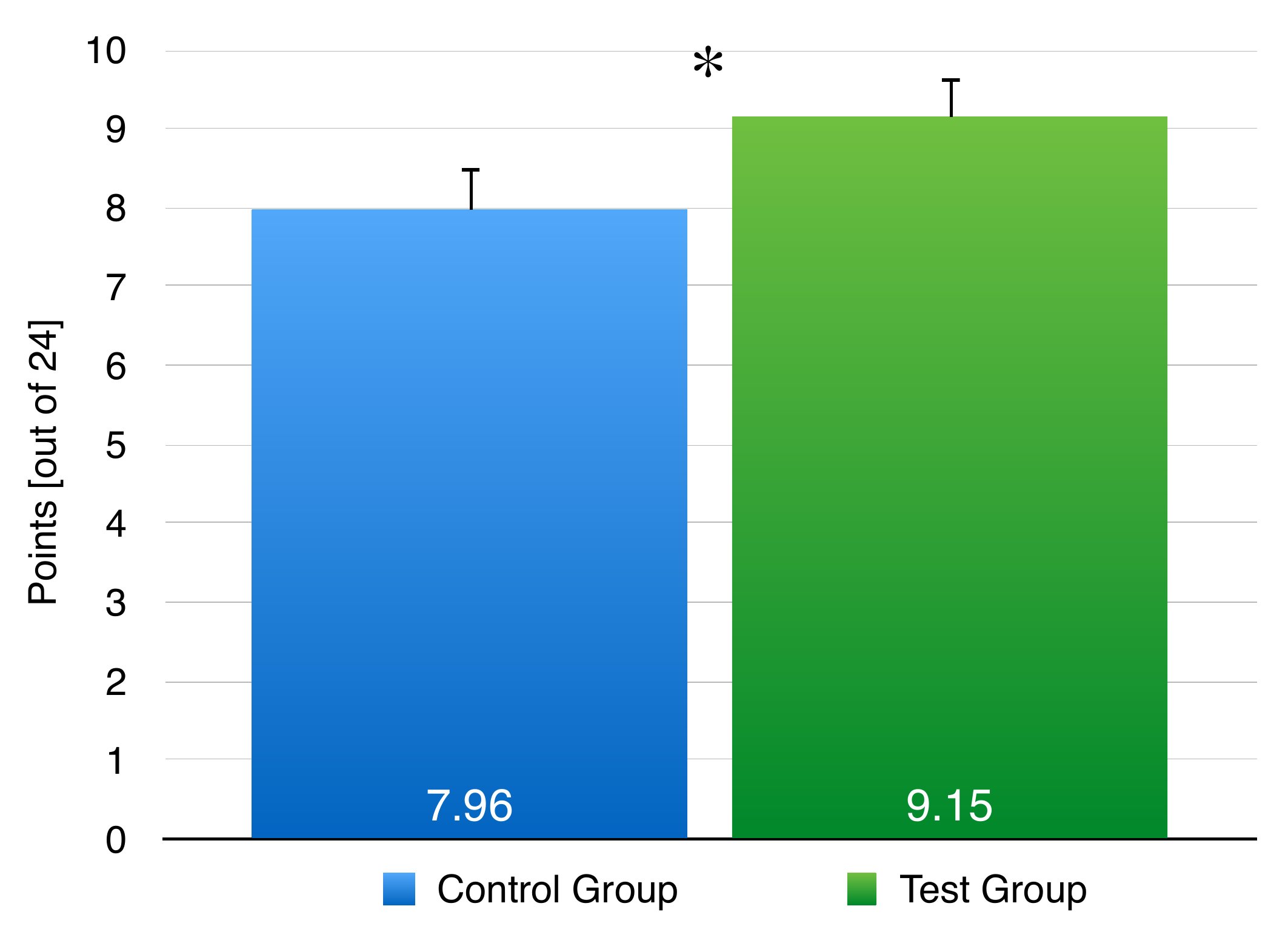}
	\end{center}
	\caption{Mean over students of the sum of scores on all of the relevant questions. A maximum of 24 points could be earned. Error bars display one standard error of the mean. The asterisk $(*)$ indicates that the difference in mean scores is statistically significant at the $\alpha=0.10$ level.}
	\label{overall}
\end{figure}

In order to evaluate our system, we collected scores on these four questions over multiple offerings of the ``Introduction to Artificial Intelligence'' course. Scores collected for offerings in 2007, 2008, 2010, and 2011 were produced by students who had no access to our system, as it had not yet been created. The students from these offerings acted as a control group. The automated grading system was used during offerings in 2012, 2013, and 2014, making the students enrolled during these years members of a test group. There were 113 students in the control group and 169 in the test group. The mean performance of students, as measured by the sum of scores received on all four of the relevant questions (24 points possible), is displayed in Figure~\ref{overall}.

\noindent
We performed a standard analysis of variance (ANOVA) of these data, using group and question as factors. This analysis revealed a marginally significant effect of group membership, with the group making use of our automated grading system receiving higher aggregate scores ($F(1,280) = 96.5$; $p = 0.066$). We also conducted planned two-tailed t-tests for each of the four relevant final examination questions, assessing the impact of our automated grading system on student performance on each question type.

The first question involved a simple translation of an English sentence into first-order logic. For example, students might be asked to translate the sentence:  ``A block can never be on top of another block that is smaller than it.'' For this final examination question, we found a marginally significant benefit of use of our system ($t(280) = 1.929$; $p = 0.055$).

The second question addressed the representation of uniqueness. An example sentence would be: ``There is exactly one block that is smaller than all of the others.'' Use of our system did not reliably influence performance on this question ($t(280) = 0.304$; $p = 0.761$).

The third question asked students to provide a definition for a predicate. Often, the question demanded the formulation of a recursive definition. For example, students might be asked to provide a definition for a simple blocks-world predicate like $\mbox{\it Above}(x,y)$ when given a predicate like $\mbox{\it On}(x,y)$. A sample solution would be:
\begin{displaymath}
	\forall x \forall y~Above(x,y) \Leftrightarrow On(x,y) \vee
        \left( \exists z~On(x,z) \wedge Above(z,y) \right)
\end{displaymath}
Use of our automated grading system produced a reliable increase in scores for this question ($t(280) = 2.077$; $p = 0.039$).

\begin{figure}[!ht]
	\begin{center}
		\includegraphics[scale=0.45]{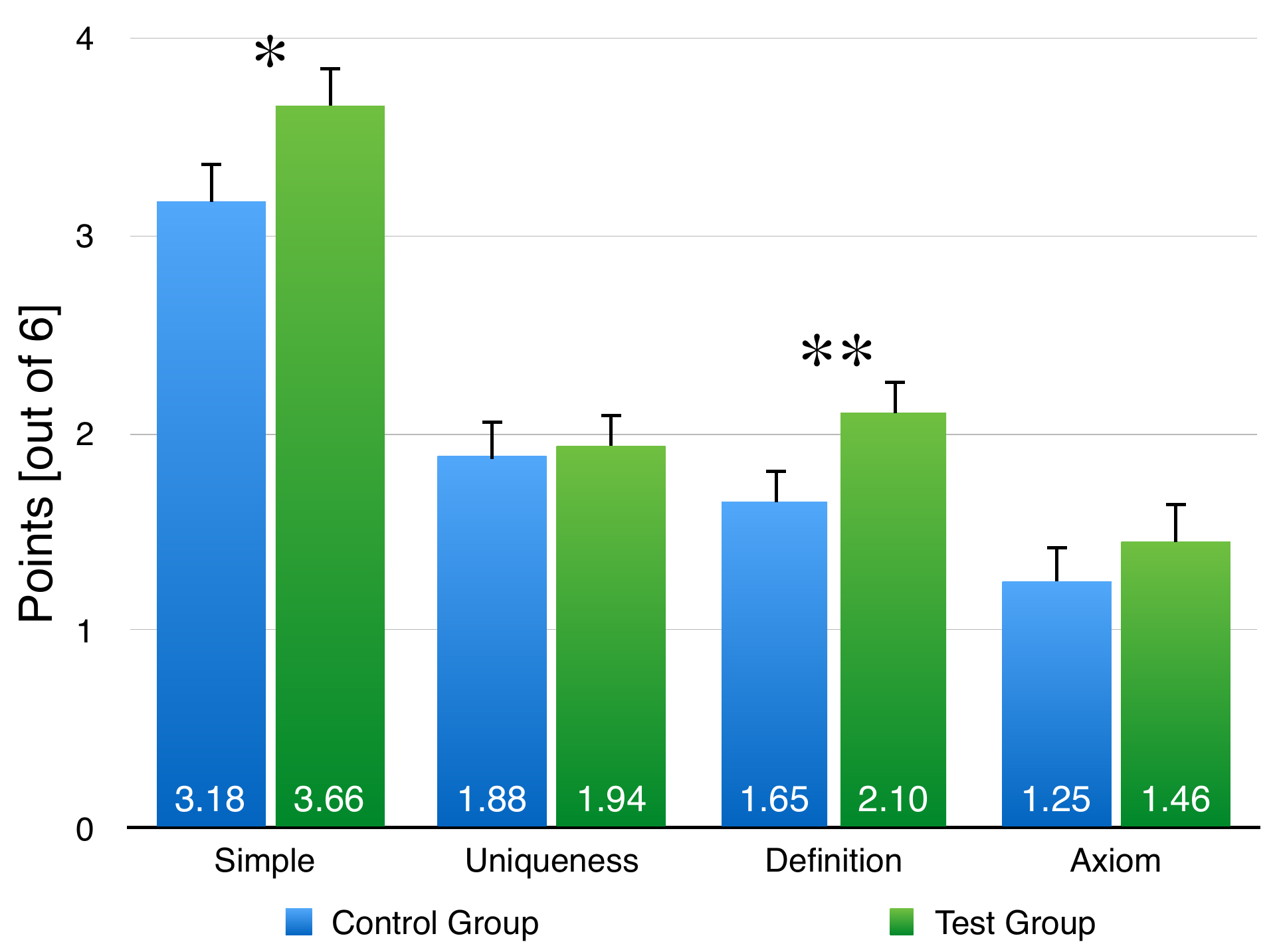}
	\end{center}
	\caption{Mean scores for each question type. The maximum possible score for each question was 6 points. Error bars display one standard error of the mean. An asterisk $(*)$ indicates that the difference in mean scores is statistically significant at the $\alpha = 0.10$ level, and a double asterisk $(**)$ marks significance at the $\alpha = 0.05$ level.}
	\label{plot}
\end{figure}

Finally, the fourth question required students to write a successor-state axiom for a given fluent using the situation calculus~\cite{russell2010}. Completing practice exercises using our system had no detectable impact on scores for this question ($t(280) = 0.856$; $p = 0.393$). The mean scores for each question are shown in Figure~\ref{plot}.

\noindent
It is worth noting that most of the exercises presented by our online system were similar to the first examination question, described above. A small number of exercises dealt with uniqueness, and there were no definition or successor-state axiom questions in the system. (Examples of definition sentences and successor-state axioms were discussed during class lectures, but the automated grading system offered no additional practice on these kinds of questions.) This observation suggests that practice on the first kind of question actually transferred to definition questions. 

\section{Future Work}
\label{future_work}

We are currently investigating a specific technical enhancement of our automated grading system to allow for the delivery of more rich feedback than a binary ``Correct/Incorrect'' assessment. This enhancement involves replacing the resolution theorem prover that is used to check for a logical equivalence between a submission and the instructor's model answer with a tableaux-based prover~\cite{Fitting:1996}. The utility of this modification stems from the fact that tableau procedures produce a counter model when the provided sentence is not valid. Such a counter model could provide useful guidance to students to help them determine the source of their misunderstandings. Consider the following example.
\\\\
Translate the sentence ``\emph{Joe does not have a lawyer}'', given the following:

\begin{center}
\begin{tabular}{l l}
	$Joe$ & A person named Joe\\
	$Lawyer$ & An occupation of being a Lawyer\\
	$Occupation(x, y)$ & person $x$ has occupation $y$\\
	$Customer(x, y)$ & person $x$ is a customer of person $y$\\
\end{tabular}
\end{center}

\noindent
One correct solution to the above exercise, call it $A$, is: 
\begin{displaymath}
\neg (\exists x~Occupation(x, Lawyer) \wedge Customer(Joe, x))	
\end{displaymath}

\noindent
\ldots and a typical incorrect solution generated by students, call it $S$, is \ldots
\begin{displaymath}
\exists x~Occupation(x, Lawyer) \wedge \neg Customer(Joe, x)	
\end{displaymath}

\noindent
Formula $S$ states that, ``There is at least one lawyer that Joe is not a customer of.'' This sentence has a different meaning than that communicated by the sentence that students were asked to translate. Testing $A \Leftrightarrow S$ for validity using the \textit{ProofTools}~\cite{prooftools} tableau prover produces the following counter model: 
\begin{displaymath}
	Occupation(B, Lawyer),~\neg Customer(Joe, B), 
\end{displaymath}
\begin{displaymath}
	Occupation(C, Lawyer),~Customer(Joe, C)
\end{displaymath}

\noindent
This counter model is similar to what a human instructor may point out to a student who has submitted $S$ as the solution for the above exercise. (``Can you see how your solution would be true if there was one lawyer, B, who was not hired by Joe, but there was another lawyer, C, who Joe retained? But the original English sentence states that Joe has no lawyer, at all.'')

Another example is: \emph{``Joe is an actor but he also has another job.''} A correct solution for this exercise is:
\begin{displaymath}
	Occupation(Joe, Actor) \wedge \exists x~Occupation(Joe, x) \wedge \neg \left( x = Actor \right)
\end{displaymath}

\noindent
Students would often forget to specify that $x$ has to be different from $Actor$ in this logical sentence. The following counter model is produced by \textit{ProofTools} when such an error is made:
\begin{displaymath}
	Occupation(Joe, Actor),~Occupation(Joe, C),~C = Actor
\end{displaymath}
\noindent
Once again, the information provided in the counter model should help a struggling student discover the error that had been made. These examples illustrate the potential for substantial improvement to the automated grading procedure that we have used.

We are also investigating strategies for improving the quality of feedback generated by automated grading systems for computer programming exercises, typically involving writing code in Java or C++~\cite{kyrilovnoelle2014}. We have reason to believe that instant feedback of a binary ``Correct/Incorrect'' nature, for exercises of this kind, has negative effects on students, as it can be demotivating for beginners and can promote cheating. Like the tableaux-based counter model feedback discussed above, we are seeking ways to automatically provide more rich and informative feedback to students completing online exercises.

\section{Conclusion}
\label{conclusion}

Undergraduate students of artificial intelligence regularly experience difficulties with knowledge representation exercises. This is evident in the low mean scores on relevant final examination questions that we have reported for our AI class. Reasons for this difficulty may include the limited amount of lecture hours devoted to first-order logic knowledge representation examples and students' relative inexperience with the subject matter. 

We have developed an automated assessment system for exercises in which students are asked to translate English sentences into first-order logic. Our objective was to give students more practice with knowledge representation, which would lead to improved performance on final examination questions. We deployed the system in our AI class, and it has been in use over the last three instantiations of the course. An analysis of examination scores shows a statistically significant improvement in the performance of students who have used our system.

In attempting to address a major shortcoming of the system, namely that it provides binary feedback, we have started investigating tableaux-based provers to power our automated grading system. This new approach has drawn our interest because tableaux procedures produce counter models for incorrect student submissions, and information from these counter models can be used to provide students with more elaborated and more meaningful feedback. Initial explorations of this idea show promise.

\section*{Acknowledgments}
The authors would like to thank Valentin Goranko for providing constructive feedback on early drafts of the paper. Thanks are also due to the anonymous reviewers who provided helpful suggestions for improving the paper.

\newpage
\thispagestyle{empty}
{\ }

\end{document}